# Tailoring Sub-micrometer Periodic Surface Structures via Ultrashort Pulsed Direct Laser Interference Patterning

F. Fraggelakis [1*], G. D. Tsibidis [1♣] and E. Stratakis [1,2♦]

[1] *Institute of Electronic Structure and Laser (IESL), Foundation for Research and Technology (FORTH), N. Plastira 100, Vassilika Vouton, 70013, Heraklion, Crete, Greece*

[2] *Department of Physics, University of Crete, 71003 Heraklion, Greece*

Direct Laser Interference Patterning (DLIP) with ultrashort laser pulses (ULP) represents a precise and fast technique to produce tailored periodic sub-micrometer structures on various materials. In this work, an experimental and theoretical approach is presented to investigate the previously unexplored fundamental mechanisms for the formation of unprecedented laser-induced topographies on stainless steel following proper combinations of DLIP with ULP. The combined spatial and temporal shaping of the pulse increase the level of control over the structure whilst it brings new insights in the structure formation process. DLIP is aimed to determine the initial conditions of the laser-matter interaction by defining an ablated region while double ULP are used to control the reorganization of the self-assembled laser induced sub-micrometer sized structures by exploiting the interplay of different absorption and excitation levels coupled with the melt hydrodynamics induced by the first of the double pulses. A multiscale physical model is presented to correlate the interference period, polarization orientation and number of incident pulses with the induced morphologies. Special emphasis is given to electron excitation, relaxation processes and hydrodynamical effects that are crucial to the production of complex morphologies. Results are expected to derive new knowledge of laser-matter interaction in combined DLIP and ULP conditions and enable enhanced fabrication capabilities of complex hierarchical sub-micrometer sized structures for a variety of applications.

## I. INTRODUCTION

Laser surface processing has emerged as a fast, chemical-free technology for surface functionalization. In particular, the use of femtosecond (fs) pulsed laser sources for material processing and associated laser driven physical phenomena have received considerable attention due to the important technological applications [1-6]. These abundant applications require a precise knowledge of the fundamentals of laser interaction with the target material for enhanced controllability of the resulting modification of the irradiated target. The physical mechanisms that lead to surface modification have been explored both theoretically and experimentally [7-18].

Various types of surface structures generated by laser pulses and more specifically, the so-called laser-induced periodic surface structures (LIPSS) on solids have been studied extensively [1, 7, 8, 14, 16, 17, 19-25]. A thorough knowledge of the fundamental mechanisms that lead to the LIPSS formation provides the possibility of generating numerous and unique surface biomimetic structures [3, 26-31] with multi-dimensional symmetry and complexity, exhibiting a broad range of sizes and spatial periodicities for a range of applications, including microfluidics [2, 32], tribology [33-35], tissue engineering [32, 36] and advanced optics [26, 31, 37]. The main technique for laser-based surface texturing is through a single step process with spatially concentrated focused laser beam (on time scales shorter than the electron-phonon relaxation time) in which an inhomogeneous energy deposition leads to self-assembly and LIPSS formation. The features of the induced periodic structures are related to the laser parameters while a series of multiscale phenomena such as energy absorption, excitation, relaxation phenomena, phase transitions and melt fluid dynamics upon resolidification determine the final relief.

Direct Laser Interference Patterning (DLIP) constitutes an alternative and high-resolution method for producing micro-nanoscale large area surface structures on metallic, semiconducting and polymeric targets. In contrast to previous techniques, this method is based on the production of a periodic interference pattern through the use of a series of overlapping coherent beams [38, 39]. The direct material removal through ablation prescribes a predefined surface topography that can be controlled by the angle of incidence of the constituent beams. All these aspects have demonstrated that DLIP is capable to offer great flexibility in the production of complex hierarchical functional structures for potential applications. For example, surface functionalities that include improved wetting properties, enhanced tribological efficiency and bacteria repellency have been already demonstrated by means of DLIP in previous works [40-43]. Furthermore, it has been shown that this technique can be employed for decorative applications by forming structural colors on the material surface, for various materials including steels and polymers [40-43]).

Nevertheless, despite the extensive research that has been conducted towards investigating the features of the surface patterns textured with DLIP, to the best of our knowledge, a detailed analysis of the physical processes that account for the structure formation due to DLIP have yet to be explored. In a recent study [44], a thermal model was introduced to present energy absorption, electron excitation and relaxation processes to calculate the thermal response of metallic materials following irradiation of a flat surface with a *single* DLIP pulse. One characteristic, though, that influences the thermal response of the material is the amount of the absorbed energy which is also closely related to the electron excitation levels and dynamics and the optical parameters of the irradiated solid. On the other hand, it is known that the generation of sub-micrometer periodic structures as a result of irradiation with laser femtosecond pulses requires exposure to many pulses (see [1] and references therein); thus, an accurate model needs to take into account the influence of varying corrugation on the energy absorption [7, 45]. In another work, a thermal model was also used to calculate the ablation depths by considering an estimation of the temporal change of the optical properties assuming an electron temperature ($T_e$) dependent variation of the reflectivity and the absorption coefficient [46]. However, the $T_e$ dependence of the optical parameters was computed by using approximate expressions for copper which are expected to be: (i) inaccurate at high temperatures (if ablation conditions are assumed) [47], (ii) inapplicable to other materials. Moreover, the thermal models used in the above studies do not take into account ablation; it is noted that consideration of the presence of a very hot part of the material throughout the relaxation processes yields overestimated values for the thermal response of the lattice.

On the other hand, it is known that to provide a consistent approach of the description of the physical mechanisms that lead to surface patterning, a multiscale approach is required, including the incorporation of processes related to mass removal (i.e. ablation), phase change and fluid hydrodynamic movement. Experimental evidence demonstrated the crucial role of the microfluidic motion on 2D-LIPSS formation [48]. Surface texturing is a multipulse process and therefore, the fundamentals of the formation of the various structures require a thorough knowledge of both intra- and inter-pulse physical effects [7, 16] as well as a precise evaluation of the absorbed laser energy to accurately describe the generation of laser induced structures. Therefore, the elucidation of the aforementioned issues is of paramount importance not only to reveal the underlying physical mechanisms of laser-matter interactions but also to improve material processing. Another aspect that, yet, has not been investigated either theoretically or experimentally is the combined action of DLIP and Double Pulse (DP) irradiation. In previous works, temporally shaped femtosecond laser pulses have been employed to control thermal effects and improve micro/nanoscale material processing. The DP approach tailors surface patterns by controlling the spatial distribution of heat [49, 50]. An interesting question is whether a combined action of a DLIP and DP technique could present a novel methodology towards controlling further the ultrafast processes that lead to surface patterning. More specifically, in a complementary way, DLIP could be employed to set the initial conditions of the structure formation process while the DP could allow a control over the evolution of the microfluidic surface reorganization.

Therefore, to fully understand the surface patterning mechanisms through the combined DLIP and DP technique, an experimental and theoretical approach is presented in this work to illustrate the plethora of the underlying complex physical processes. Special emphasis is given on the description of: (i) the energy absorption through the use of data obtained from Density Functional Theory simulations and the energy absorption of a liquid material assuming a dynamical change of the optical parameters [51, 52], (ii) electron excitation and dynamics, (iii) mass removal and (iv) hydrodynamical phenomena that determine the surface topography following a multi-pulse process. To account for the capability to intervene in the material reorganization process, a detailed description of the fundamental mechanisms that determine the surface topography is investigated; to this end, both single (SP) and temporally delayed DLIP double pulses (DP) are used to estimate the influence of different absorption and excitation levels when the second of the DP irradiates a material in molten phase. To illustrate the role of the periodicities of the interference patterns on the surface features (i.e. frequencies of induced structures, height, ablated depth, complexity of sub-micrometer periodic structures, etc.) a DLIP technique with variable induced periodicity $\Lambda_{LIPSS}$ is used while a multiscale model is presented that incorporates the influence of electrodynamical effects (i.e. excitation of Surface Plasmon Polaritons, SPP) on the formation of the surface topography. To test the validity of the theoretical model in laser conditions that lead to novel morphologies that has not been previously investigated, a two and four beam DLIP-based irradiation with SP and DP ultrashort pulses is also experimentally explored. Observations indicate that (i) pulse separation, (ii) number of beams of the DLIP, and (iii) angle of incidence of constituent pulses are capable to fabricate an abundance of not previously produced morphologies.

To this end, the present work is organised as follows: in Section II, the experimental protocol is illustrated to describe the DLIP-based set up that was developed to control the production of various morphologies of different feature sizes and complexities following irradiation of stainless steel with femtosecond pulses. While the periodicities of the DLIP are taken to be of the size of the laser wavelength ($\lambda_L$~1026 nm) or 4-6 times larger than $\lambda_L$, the employment of ultrashort pulsed lasers lead to the generation of sub-micrometer periodic structures. In Section III, a detailed multiscale theoretical framework is presented to describe the physical mechanisms that account for production of the induced surface structures in various conditions. A systematic analysis of the results is illustrated in Section IV while concluding remarks follow in Section V.

## II. EXPERIMENTAL PROTOCOL

Experiments are performed utilizing 1D and 2D DLIP combined with SP and DP irradiation. Several techniques have been introduced in order to realize DLIP, including the use of a grating, a prism and a lens to combine the laser beams, as well as a Spatial Light Modulator (SLM) [53]. Nevertheless, femtosecond cannot be used with some of these configurations. The lack of spatial coherence, in particular, emerges as the main drawback due to the limited pulse length and the unavoidable difference of the optical paths of the interfering beams. Furthermore the angle of the incident beam limits the interference volume in a limited region of the irradiated area [38]. Both issues can be resolved by employing a grating to divide the laser beams complemented with an appropriate imaging system [54]. In this work, instead of a fixed grating, an SLM module is employed as a variable grating in order to control the angle and the number of the incident laser beams.

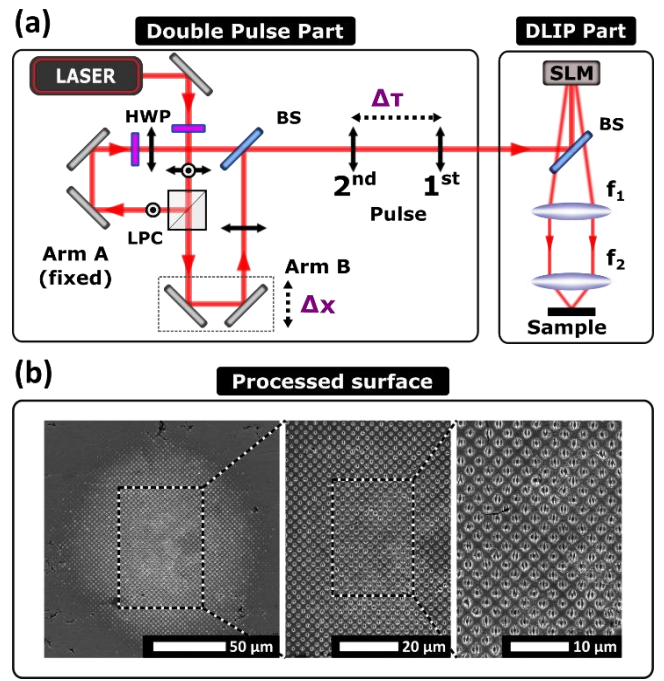

FIG. 1. (a) Experimental setup of combined DLIP and DP. Abbreviations: Half-waveplate (HWP), Linear polarizing cube (LPC), Beamsplitter (BS), Spatial Light Modulator (SLM), focusing lens (f) (b) Surface processed with 4 beams with $\theta_n = 19 \pm 0.5$ °, $NP = 50$ and 42 μJ per pulse.

To investigate the role of DLIP size and the impact of a delayed pulse in the features of the induced surface pattern, a Pharos laser source emitting femtosecond pulses of pulse duration $\tau_p \cong 170$ fs at $\lambda_L = 1026$ nm is employed. The setup is divided into two parts, the DP part and the DLIP part as indicated Fig.1. The generation of DP occurs due to a modified interferometer shown in Fig.1a. The laser beam is then guided to the DLIP part where multiple beams are generated and recombined. A programmable SLM is utilized as a tunable diffraction grating. Phase masks are applied to generate two or four beams diverging in different angles. Two focusing lenses $f_1 = 400$ mm and $f_2 = 30$ mm placed on the appropriate distances are used to recombine the beams on the sample. Owing to this setup [54] it is possible to overcome the issue of coherence of femtosecond pulses that prevents successful generation of DLIP [38] and acquire interference pattern throughout the whole area of the irradiated spot. The surface pattern resulting from the femtosecond DLIP irradiation of stainless steel, generated by four beams and having incident angle of $\theta_n = 19 \pm 0.5$ ° is illustrated in Fig.1b.

For all experiments, a commercially available 316 stainless steel has been used. The energy per pulse was 42 μJ or 50 μJ and the total number of pulses (*NP*) incident to the surface was varied from *NP*=10 to *NP*=500. Two and four fs beams were employed to generate DLIP patterns having 1D and 2D symmetry, respectively. Periods of DLIP structures ($\Lambda_{DLIP}$) that were used were either comparable to the laser wavelength or ~$5\lambda_L$. For the case of 1D DLIP pattern, the orientation of the laser polarization was perpendicular to the DLIP pattern to generate LIPSS parallel to the DLIP groove. The experimental process was divided in to two parts, related to the combination of DLIP with SPI and DPI respectively. Images of the processed surfaces were acquired via Scanning Electron Microscopy (SEM) microscopy and a Fast Fourier Transformation (FFT) was performed to calculate the periodicities of the induced structures through the use of the open source software Gwyddion.

### III. THEORY

**a. DLIP**

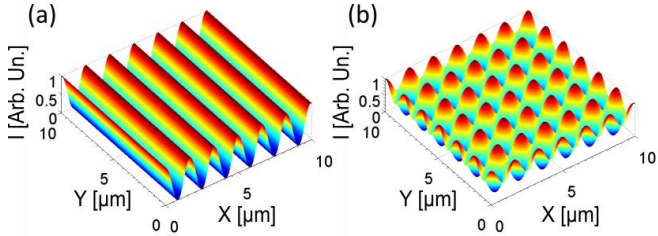

FIG. 2. Normalised intensity distribution for (a) two- and (b) four-beam interference. Periodicities equal to *P*= 1650 nm along *x*-axis (for (a)) and *x*- and *y*-axis (for (b)) have been selected.

The physical mechanism on which DLIP interference is based is the superposition of the electric fields of at least two coherent laser beams according to the following scheme (on the surface $z_s$ of the material and at position defined by coordinates (*x,y*)) [38, 39]

$$\vec{E}_{total}(z_s) = \sum_{n=1}^{N} \vec{E}_{n0} e^{-i(\omega t - \vec{k} \cdot \vec{r})} = \sum_{n=1}^{N} \vec{E}_{n0} e^{-i(\omega t - 2\pi/\lambda_L \sin(\theta_n/2)[x\cos(\beta_n) + y\sin(\beta_n)])} \quad (1)$$

where $|\vec{E}_{n0}|$ is the amplitude of the electric field of the *n*-th beam while $\vec{E}_{n0}$ includes the polarisation direction, $\omega$ stands for the angular frequency, *t* is the time. Each laser beam irradiates the material at an incident angle with the vertical axis equal to $\theta_n/2$ and azimuthal angle $\beta_n$. The total spatial intensity distribution is, then, provided by the expression $I_0(z_s) = c\varepsilon_0 |\vec{E}_{total}|^2/2$, where *c* and $\varepsilon_0$ are the speed of light and dielectric vacuum permittivity, respectively. For interference with *two* ($\beta_1=\beta_2=0$, and $\theta_1 = \theta_2 = \theta$) and *four* beams ($\beta_1=\beta_2=0$, $\beta_3=\beta_4=\pi/2$, $\theta_1 = \theta_2 = \theta_3 = \theta_4 = \theta$) the following total intensities are $I_0^{(2)}, I_0^{(4)}$ produced, respectively

$$I_0^{(2)}(z_s) \sim I_1 \left[ \cos\left(\frac{4\pi}{\lambda_L} x \sin\left(\frac{\theta}{2}\right)\right) + 1 \right] e^{-4\log(2)\left(\frac{x^2+y^2}{R_0^2}\right)}$$

$$I_0^{(4)}(z_s) \sim I_1 \left\{ \left[ \cos\left(\frac{4\pi}{\lambda_L} x \sin\left(\frac{\theta}{2}\right)\right) + \cos\left(\frac{4\pi}{\lambda_L} y \sin\left(\frac{\theta}{2}\right)\right) + 2 \right] \right\}$$
$$\times e^{-4\log(2)\left(\frac{x^2+y^2}{R_0^2}\right)} \quad (2)$$

where $I_1$ is the intensity of each of the constituent laser beams of the DLIP. It is noted that energy deposition is considered assuming Gaussian beams of FWHM equal to $R_0$.

It is evident (Eqs.2) that the choice of $\theta_n$ can be used to define the periodicities of the interference pattern. More specifically, for two- and four-beam DLIP, a sinusoidal (of periodicity equal to $\Lambda_{DLIP} = \frac{\lambda_L}{\left(2\sin\left(\frac{\theta}{2}\right)\right)}$ along *x*-axis) or dot-type intensity distribution (of periodicity equal to $\Lambda_{DLIP} = \frac{\lambda_L}{\left(2\sin\left(\frac{\theta}{2}\right)\right)}$ along *x*-axis and *y*-axis) is derived as illustrated in Fig.2.

**b. Electron excitation and relaxation process**

The two-temperature model (TTM) constitutes the standard theoretical framework to investigate laser-matter interaction upon femtosecond laser irradiation [55]. A 3D-TTM is implemented by the following set of coupled differential equations that describe the absorption of optical radiation by the electrons and the energy transfer between the electron and lattice subsystems

$$\begin{aligned} C_e \frac{\partial T_e}{\partial t} &= \vec{\nabla}(k_e \vec{\nabla} T_e) - g(T_e - T_L) + W \\ C_L \frac{\partial T_L}{\partial t} &= \vec{\nabla}(k_L \vec{\nabla} T_L) + g(T_e - T_L) \end{aligned} \quad (3)$$

where $C_e$ and $C_L$ stand for the heat capacities of the electron and lattice subsystems, respectively while $T_e$ and $T_L$ are the temperatures of the two systems. On the other hand, $k_e$ ($k_L \sim 0.01\, k_e$) correspond to the electron (lattice) conductivity, *g* is the electron-phonon coupling parameter while *W* corresponds to the absorbed laser power density which is provided through the following expressions (taken from Eqs.2) [7, 8, 10, 56-59]

$$-\frac{\partial I(t,x,y,z)}{\partial z} = \alpha(t,x,y,z) I(t,x,y,z) = W(t,x,y,z) \quad (4)$$

$$I(t,x,y,z_s(x,y)) = \left(1 - R(t,x,y,z_s(x,y))\right) I_1(t,x,y,z_s(x,y)) \quad (5)$$

$$I_1(t,x,y,z_s(x,y)) = \frac{2\sqrt{\log(2)}}{\sqrt{\pi}\tau_p} \frac{F}{2} \left[ e^{-4\log(2)\left(\frac{t-3\tau_p}{\tau_p}\right)^2} + e^{-4\log(2)\left(\frac{t-3\tau_p-\tau_d}{\tau_p}\right)^2} \right] \quad (6)$$

In the above expressions, $R$ and $α$ stands for the reflectivity and the absorption coefficient of the material, $2F$ is the fluence of the DLIP pulse (i.e. each of the constituent pulses of the DLIP pulse is assumed to have fluence equal to $F$) and $τ_d$ is the temporal delay between the two pulses in the DP experiment (i.e. $τ_d = 0$, for a single pulse). Furthermore, $I(t,x,y,z_s(x,y))$ corresponds to the value of the intensity on the surface of the material. As a Cartesian coordinate system is used, the position of the surface in the vertical axis, $z_s$, varies with $(x,y)$ as the surface morphology changes locally due to the corrugated profile. More specifically, for $NP=1$ (i.e. flat surface), the position of the surface is at $z_s = 0$. By contrast, for $NP>1$, as the surface morphology changes, $z_s$ becomes dependent on the $(x,y)$ position. To compute the intensity $I$ at positions below the surface, the following expression is used recursively

$$I(x,y,z) = I(x,y,z-dz) - \frac{\partial I(t,x,y,z)}{\partial z} dz \qquad \text{for } z > z_s \quad (7)$$

while at $z=z_s$, $I$ is provided from Eq.5. It is noted that $dz$ represents infinitesimally small increments of $z$ and it is used to compute the attenuation of the laser energy inside the irradiated volume. With respect to the energy that is absorbed from the surface of the material, Eq.5 is used (see also Refs. [7, 8, 10, 56-59]; in other reports, to account for the difference in the optical response and the the role of inhomogeneous depth profile, a multilayer was used to calculate the transient reflection coefficient [60]).

Various methodologies have been proposed to calculate the thermophysical properties of the material (i.e. electron heat capacity, conductivity, electron-phonon coupling constant); among the most accurate are those that involve a computation of the density of states (DOS) for various energies below and above the Fermi energy

**Table I. Simulation parameters chosen for 100Cr6 steel [22]**

| Parameter | Value |
|---|---|
| $A$ [s$^{-1}$ K$^{-2}$] | 0.98×10$^7$ [22] |
| $B$ [s$^{-1}$ K$^{-1}$] | 2.8×10$^{11}$ [22] |
| $k_{e0}$ [Wm$^{-1}$K$^{-1}$] | 46.6 [61] |
| $C_L$ [J kg$^{-1}$K$^{-1}$] | 475 [61] |
| $C_L^{(m)}$ [J kg$^{-1}$K$^{-1}$] | 748 [62] |
| $T_{melt}$ [K] | 1811 [63] |
| $T_{cr}$ [K] | 8500 [64] |
| $T_{boiling}$ | 3100 [64] |
| $ρ_0$ [kg m$^{-3}$] | 6900 [62] |
| $μ$ [Pa s] | 0.016 [65] |
| $σ$ [Nm$^{-1}$] | 1.93-1.73×10$^{-4}$($T_L$-$T_{melt}$)K$^{-1}$ [66] |
| $L_v$ [J g$^{-1}$] | 6088 [62] |
| $L_m$ [J g$^{-1}$] | 276 [62] |
| $R_0$ [μm] | 145 |
| $τ_p$ [fs] | 170 |
| $τ_d$ [ps] | 500 |

[67]. More specifically, the effect of the thermal excitation of electrons on properties such as the electron-phonon coupling and electron heat capacity can be determined through the characteristics of the electron DOS [67]. Nevertheless, while such information exists for a large number of known metals [67], there is a lack of knowledge of these parameters for materials used for industrial applications such as 316 stainless steel which is the material used in this work. A rigorous approach would be to use first principles and derive, firstly, the DOS for this material by using relevant software, density functional theory and experimental data [68] and, secondly, produce an estimate for those parameters. Herein, a simplified approach is followed in which an approximation is performed based on the fact that iron (Fe) is the main ingredient of the stainless steel [61]. The employment of the thermophysical properties based on the fitting of data for Fe does not differ significantly in various types of stainless steel. Indeed, recent results indicate that the temperature dependent electron heat capacity of a steel alloy is not substantially different from that predicted for Fe [69]. Similarly, previous calculations indicate that a more rigorous computation of the electron-phonon coupling is not anticipated to produce substantially different morphological results [22]. Therefore, the (electron) temperature dependent heat capacity $C_e$ and electron-phonon coupling strength $g$ of Fe are computed using a polynomial fitting of calculated values [67]. It is also noted that other thermophysical parameters of the material used in this work are approximated with results of 100Cr6 stainless steel (i.e. various types of steel do not show significantly different thermophysical properties ). For example, the heat conductivity is calculated from the expression $k_e = k_{e0} \frac{BT_e}{A(T_e)^2 + BT_L}$ (the parameters $A$ and $B$ have been obtained from variable angle spectral ellipsometric measurements of the refractive index and the extinction coefficient of the polished 100Cr6 steel at various wavelengths [22]). A summary of the values of the parameters used in the simulations are shown in Table I.

**c. Optical parameters**

It is well known that the transient variation of the dielectric parameter through the $T_e$ dependence of the electron relaxation time leads to a change of the optical properties of the material during the irradiation time that needs to be evaluated since it influences the absorbed energy. Therefore, the, usually, constant value of the optical parameters that is assumed in simulations for metals is a rather crude approximation. A more complete approach is necessary that will involve a rigorous consideration of changes in the optical properties during the duration of the pulse (see discussion in Ref.[58]). Given that effects due to DP are also investigated, and since conditions are explored in which material experiences a phase transition before the delayed pulse irradiates it, a two tiered approach is followed in the current study: (i) results from DFT calculations are used to express the dynamic change of the optical parameters of the irradiated material [51] (see [70]), (ii) results for the reflectivity values of Fe in liquid phase are used to describe the energy absorption when the delayed pulse irradiates the material (reflectivity varies from 20% to 60% for $T_L$ values between $2T_{melt}$ and $T_{melt}$ for $\lambda_L$~1.03 μm [52]).

**d. Ablation**

To simulate ablation, a previously proposed process to model mass removal is used. More specifically, a solid material that is subjected to ultrashort pulsed laser heating at sufficiently high fluences undergoes a phase transition to a superheated liquid with temperatures that exceed $0.90T_{cr}$ ($T_{cr}$ being the thermodynamic critical temperature, $T_{cr}$(Fe) = 8500 K) [71]. According to Kelly and Miotello [71], melted material at and beneath the irradiated surface is unable to boil, as the timescale does not permit heterogeneous nucleation. A subsequent homogeneous nucleation of bubbles leads to a rapid transition of the superheated liquid to a mixture of vapor and liquid droplets that are ejected from the bulk material (a process referred to as phase explosion). This is proposed as a material removal mechanism and it is assumed that phase explosion occurs when the lattice temperature is equal or greater than $0.90T_{cr}$ [7, 22, 45, 71-74].

**e. Hydrodynamical effects**

To model a surface modification following irradiation with femtosecond laser pulses, it is assumed that the laser conditions are sufficiently high to result in a phase transition from solid to liquid phase and upon resolidification a surface relief is induced. The melting point of stainless is taken as the threshold for a phase transition from solid to liquid while the $T_{melt}$ isothermal is considered as the criterion for resolidification (i.e. when $T_L$ drops below $T_{melt}$ resolidification starts). The movement of a material in the molten phase is given by the following Navier-Stokes equations (NSE) which describes the dynamics of an uncompressible fluid [75]

$$\rho_0 \left( \frac{\partial \vec{u}}{\partial t} + \vec{u} \cdot \vec{\nabla} \vec{u} \right) = \vec{\nabla} \cdot \left( -P + \mu(\vec{\nabla}\vec{u}) + \mu(\vec{\nabla}\vec{u})^T \right) \quad (8)$$

where $\rho_0$ and $\mu$ stand for the density and viscosity of molten stainless steel, while $P$ and $\vec{u}$ are the pressure and velocity of the fluid. The fluid is considered to be an incompressible fluid (i.e. $\vec{\nabla} \cdot \vec{u} = 0$).

In regard to the pressure, there are two terms that require special treatment:
- the **recoil pressure** which is related to the lattice temperature of the surface of the material through the equation [76, 77]

$$P_r = 0.54 P_0 exp\left( L_v \frac{T_L^{(S)} - T_{boiling}}{R_G T_L^{(S)} T_{boiling}} \right) \quad (9)$$

where $P_0$ is the atmospheric pressure (i.e. equal to $10^5$ Pa [78]), $L_v$ is the latent heat of evaporation of the liquid, $R_G$ is the universal gas constant, $T_{boiling}$ stands for the boiling temperature for iron and $T_L^{(S)}$ corresponds to the surface temperature. When vapour is ejected, it creates a back (recoil) pressure on the liquid free surface which in turn pushes the melt away in the radial direction [7] which results in a depression of the surface. Furthermore, given the spatially modulated energy deposition on the material, a gradient of the lattice temperature is produced which is, in turn, transferred into the fluid and therefore a capillary fluid convection is produced.
- A precise estimate of the molten material behaviour requires a contribution from the **surface tension related pressure**, $P_\sigma$, which is influenced by the surface curvature and is expressed as $P_\sigma = K\sigma$, where $K$ is the free surface curvature and $\sigma$ surface tension. The calculation of the pressure associated to the surface tension requires the computation of the temporal evolution of the principal radii of surface curvature $R_1$ and $R_2$ that correspond to the convex and concave contribution, respectively [79]. Hence the total curvature is computed from the expression $K=(1/R_1 +1/R_2)$. A positive radius of the melt surface curvature corresponds to the scenario where the centre of the curvature is on the side of the melt relative to the melt surface (see Ref. for a detailed description of the simulation methodology [7]).

Pressure equilibrium on the material surface implies that the pressure $P$ in Eq.8 should outweigh the accumulative effect of $P_r + P_\sigma$. The thermocapillary boundary conditions imposed at the liquid free surface are the following

$$\frac{\partial u}{\partial z} = -\sigma/\mu \frac{\partial T_L}{\partial x} \quad \text{and} \quad \frac{\partial v}{\partial z} = -\sigma/\mu \frac{\partial T_L}{\partial y} \quad (10)$$

where $(u,v,w)$ are the components of $\vec{u}$ in Cartesian coordinates. The cartesian coordinate system indicated by $(x,y,z)$ is used to describe morphological changes compared to the initial $(x,y,z_S)$ for flat surfaces.

It is noted that a more precise evaluation of the fluid material parameters such as the surface tension, viscosity, recoil pressure and density at elevating (above the melting point and below $T_{cr}$) temperatures would allow a more realistic description of the fluid dynamics (see Ref. [7]). As such values (to the best of our knowledge) have not been reported, without loss of generality, the values stated in Table I will be used in this work.

**f. Surface plasmon excitation**

According to the Surface Plasmon Polariton (SPP)-model, the dispersion relation for the excitation of SPP is derived by the boundary conditions (continuity of the electric and magnetic fields at the interface between a metallic and dielectric material) ($\varepsilon_d = 1$) for a flat surface (number of pulses (*NP*), *NP*=1). Therefore, a requirement for a semiconductor to obey the above relation and conditions is that $Re(\varepsilon) < -1$ and the computed SPP wavelength $\Lambda_{SPP}$ is given by the expression $\Lambda = \lambda_L/Re\sqrt{\frac{\varepsilon}{\varepsilon+1}}$ [7, 56] where $\varepsilon$ stands for the dielectric parameter for irradiation in vacuum which is approximately correct for nearly flat surfaces and very small *NP* [21]. As shown in previous works [7, 80], the interference of SPP waves with the incident laser beam (only after a corrugation on the surface or a small crater has been created) leads to a periodic modulation of the absorbed energy that yields a periodic variation of the thermal and hydrodynamical properties [7]. As a result, a periodic surface pattern is produced with the formation of Low Spatial Frequency LIPSS (LSFL-SPP, in which SPP indicates that LSFL is generated from SPP) which are orientated perpendicularly to the laser beam polarisation. On the other hand, it is noted that results of the computed value of SPP and the periodic structures that are formed differ from the one computed through the above expression as enhanced corrugation has proven to yield a shift to the SPP resonance to smaller values of $\Lambda$ at increasing *NP* [14, 21, 22]. Results were also obtained for excitation of SPP for deeper gratings [81, 82]. In contrast to electrodynamics simulations, mainly, based on Finite Difference Finite Domain Schemes (FDTD) or analytical approaches used to correlate the induced periodicities with a variable corrugation as a result of increase of the irradiation dose *NP* [11, 14, 83-86], an alternative and approximating methodology has also been employed to relate the SPP wavelength with the produced maximum depth of the corrugated profile [21, 22] (i.e. which is linked with *NP*). The methodology was based on the spatial distribution of the electric field on a corrugated surface of particular periodicity and height and how continuity of the electromagnetic fields influences the features of the associated SPP. This methodology is also used in the present work. Results of the SPP wavelength as a function of *NP* is shown in Supplementary Material [70].

### g. Components of the multiscale model

The model aims to present a consistent methodology that incorporates/couples all processes which take place in various temporal scales and predict laser-based surface patterning features. Processes such as energy absorption, electron excitation, SPP excitation, electron-phonon coupling and relaxation phenomena, phase transition, melt fluid dynamics and resolification constitutes are simulated and they are parts of a multiscale model. In comparison with the state-of-the-art modelling approaches that have been used [7, 25, 84, 86, 87], the additional features the model incorporates are the following: (i) it allows the inclusion of a complex intensity spatial profile to account for the impact of irradiation with a DLIP technique (i.e. the angle of incidence of the constituent laser Gaussian beams are considered) and, finally, predicts well-ordered, novel morphologies with 1D and 2D symmetries, (ii) it incorporates the $T_e$ dependent values of the optical properties of the irradiated material material that have been derived through rigorous DFT calculations [56], (iii) it describes electron excitation and relaxation processes following DLIP and DP; in particular, special emphasis is given on the optical response of a *fluid* irradiated with ultrashort pulses (i.e. the second constituent pulse of DP irradiates a material in a liquid phase and therefore appropriate caution is required to compute the energy absorption and excitation of molten material), (iv) it includes a transient change of the irradiated region at increasing energy dose (i.e. *NP*) that is modelled by taking into account ablation conditions.

## IV. SIMULATION PROCEDURE

To solve the set of the above equations, a scheme based on finite difference method is used. A common approach followed to solve similar problems is the employment of a staggered grid finite difference method which is found to be effective in suppressing numerical oscillations. Unlike the conventional finite difference method, temperatures ($T_e$ and $T_L$), pressure ($P$) are computed at the centre of each element while time derivatives of the displacements and first-order spatial derivative terms are evaluated at locations midway between consecutive grid points. For time-dependent flows, a common technique to solve the NSE equations is the projection method and the velocity and pressure fields are calculated on a staggered grid using fully implicit formulations [88, 89]. On the other hand, the horizontal and vertical velocities are defined in the centres of the horizontal and vertical cells faces, respectively (for a more detailed analysis of the numerical simulation conditions and the methodology towards the description of fluid dynamics, see Refs. [7, 8, 21, 22, 25, 90-92]).

The hydrodynamic equations are solved in both sub-regions that contain either solid and or molten material; for the sake of simplicity, it is assumed that regions with mixed composition do not exist [7]. To include the "hydrodynamic" effect of the solid domain, material in the solid phase is modelled as an extremely viscous liquid ($\mu_{solid} = 10^5 \mu_{liquid}$), which results in velocity fields that are infinitesimally small.

At time $t = 0$, both electron and lattice temperatures are set to room temperature (300 K). Non-slipping conditions (i.e. the spatial velocity field is zero everywhere) are applied on the solid-liquid interface. Heat loss from the upper surface of target is assumed to be negligible. As a result, a zero heat flux boundary condition is set for the electron and lattice systems. Peak fluence values $F$ equal to 0.15 J/cm$^2$ (for two-beam DLIP) and 0.5 J/cm$^2$ (for four-beam DLIP) are considered in the simulations. For *NP*=1, a 2D-numerical solution is followed due to the axial symmetry of the problem. As the material is subjected to irradiation by multiple laser pulses, Eqs.1-10 are solved in a three dimensional Cartesian coordinate system and the energy absorption in subsequent irradiation (*NP*>1) is modelled by considering a ray tracing approach to compute the absorbed and reflected part in a modified profile.

The irradiated region is split into two sub-regions to accommodate solid and molten material. The temporal calculation step is adapted so that the stability Neumann condition is satisfied [93]. In regard to the material removal simulation, in each time step, lattice and carrier temperatures are computed and if lattice temperature reaches $~T_L>0.9T_{cr}$, mass removal through evaporation is assumed. In that case, the associated nodes on the mesh are eliminated and revised boundary conditions on the new surface are enforced. It is also noted that the removal of the material points is necessary in order to describe correctly the thermal process otherwise an overheating and overestimation of the thermal effects is produced.

To summarise the adjustable parameters in the model and the simulation procedure, we note that: (i) $T_e$-dependent values are taken for the optical properties (through DFT calculations), $g$ and $C_e$ (through fitting), and $k_e$ while (ii) $T_L$ -dependent values are considered for $\mu$ $P_r$ and $P_\sigma$.

## V. RESULTS AND DISCUSSION

A detailed experimental investigation has been conducted to describe how the number of DLIP beams, DLIP periodicities, polarisation of laser beams and irradiation with SP or DP influence the generated surface pattern. It is noted that the values of fluences used in this work has been selected to produce ablation effects. Therefore, as underlined in the previous sections, in addition to the electrodynamic effects, the thermal response of the irradiated material, the ablation efficiency and hydrodynamic effects are required to be evaluated in detail in order to interpret the experimental results through consistent physical mechanisms.

To demonstrate, firstly, the impact of the DP and the fact that the second constituent pulse of DP irradiates molten material, simulations results show [70] that the first of the two pulses for *NP*=1 leads to a maximum depression of the surface equal to ~24 nm (at *x*=*y*=0 where the energy deposition is highest) due to ablation; By contrast, as the second of DP irradiates a material in molten phase and given the significantly reduced reflectivity of the fluid [70], the energy which is absorbed is enhanced which subsequently leads to accumulative ablated region equal to ~34 nm. Predicted results for the size of the ablated region appear to agree with experimental data [46]. In the next sections, experimental results are presented for SP and DP for two-beam and four-beam DLIP is also discussed. To interpret surface patterning features, simulation results based on the physical model introduced in the previous section are presented.

### a. Two laser beam DLIP with $\Lambda_{DLIP} \sim \Lambda_{LIPSS}$

To explore the influence of the DLIP period on the features of the induced pattern, stainless steel surfaces were irradiated with a combination of DLIP and trains of single and DP with a DLIP period which is comparable with the laser wavelength ($\Lambda_{DLIP}$~1650 nm). Relevant experimental results are illustrated by the SEM images in Fig.3 and Fig.4, respectively.

Results indicate that for single pulses and *NP*=10, High Spatial Frequency LIPSS (HSFL) are formed with orientation *parallel* to the laser beam orientation (Fig.3a) for *NP*=10. HSFL have spatial periods significantly smaller than the irradiation wavelength and, in metals, they occur at low fluence values (i.e. close to the ablation threshold) and small *NP* [23, 94]. By contrast, a different type of LIPSS structures is produced for *NP*=50 (and $\Lambda_{DLIP}$~1662 nm as shown in Fig.3b) with orientation *perpendicular* to the laser polarization (LP) and *parallel* to the DLIP. To illustrate the features of the produced structures a cross section of the SEM image along the pattern has been obtained (Fig.3c). Although, SEM images are not capable, in principle, to provide a precise estimate of the depth of the corrugated pattern, it is assumed that the intensity profile of the image along the patterned surface can approximately illustrate the pattern shape and morphological changes. Certainly, AFM images would allow a more precise evaluation of the measured depth. The periodicity of the structures is ~471 nm, calculated through Fast Fourier Transform (FFT) analysis of SEM images of a ~30 x 30 $\mu m^2$ region) (Fig.3d,e). Shaded areas in light blue colour shows the contour of the DLIP pattern. It is evident that LIPSS are formed randomly with respect to the DLIP.

An interesting outcome related to the observed LIPSS structures for *NP*=50 is that the observed and simulated structures exhibit deep subwavelength periodicities ($<1/2\lambda_L$) that is significantly smaller than the expected values for LSFL structures. This is in contrast to the dominant LSFL-SPP mechanism presented in the previous section for LSFL formation with orientation perpendicular to LP [7, 95], while a possible explanation can be due to electrodynamic effects (i.e. second harmonic generation [96], near fields [95], cylindrical waves [86]) a modelling approach based on the theoretical framework presented in this work is employed to further investigate the structure formation. Simulations indicate that an ablated region is, firstly, produced while fluid transport further directs the molten material movement (in Fig.3f, temperature profile is shown at time *t*=450 ps) to determine the surface relief. Due to the small size of the induced DLIP crater (i.e. diameter is smaller than ~900 nm), SPP excitation modes and coupling with the incident beam that can lead to periodic energy modulation inside the crater and formation of LSFL-SPP structures is not possible to occur. Therefore, in this case, solely the hydrodynamical response of the material can account for surface modification. According to the model, repetitive irradiation leads to corrugation profile inside the groove Fig.3g (for *NP*=10) with size comparable to the observed LIPSS (*blue* line in Fig.3h). The difference between the experimental observations and structure type in *NP*=10 and *NP*=50 can be attributed to the energy values that are not sufficiently high for *NP*=10. A similar conclusion can be deduced to explain the discrepancy of the surface pattern profiles for *NP*=10 between the experimental results and simulations that leads to HSFL structure formation. Thus, more investigation is required to describe the formation of the HSFL and the transition from HSFL to low spatial frequency LIPSS.

On the other hand, it is evident that the enhanced hydrodynamical effects which are developed (Fig.3f) lead to a deeper corrugated profile with a small peak inside the ablated region at a distance equal to ~0.22 µm from the DLIP highest position (Fig.3h) and ~ 0.53 µm away from another similar peak inside the crater. Moreover, the final profile along two DLIP periods appears to agree with the pattern shape obtained from experiments. The proposed hydrodynamic process appears to provide a consistent description. To illustrate the dynamic process that leads to the formation of the surface pattern with increasing $NP$, simulations results are presented in (Fig.3h) and in the Supplementary Material [70]). More specifically, the depth profile along the same place (*white* dashed line in Fig.3g) is illustrated in Fig.3h for $NP$=5, 10, 15, 20. It is noted that while the phase transition occurs in the ps-scale completion of resolidification process requires a few ns in the fluence range considered in the experimental process. The time required for resolidification has been confirmed in previous reports [7, 97-100].

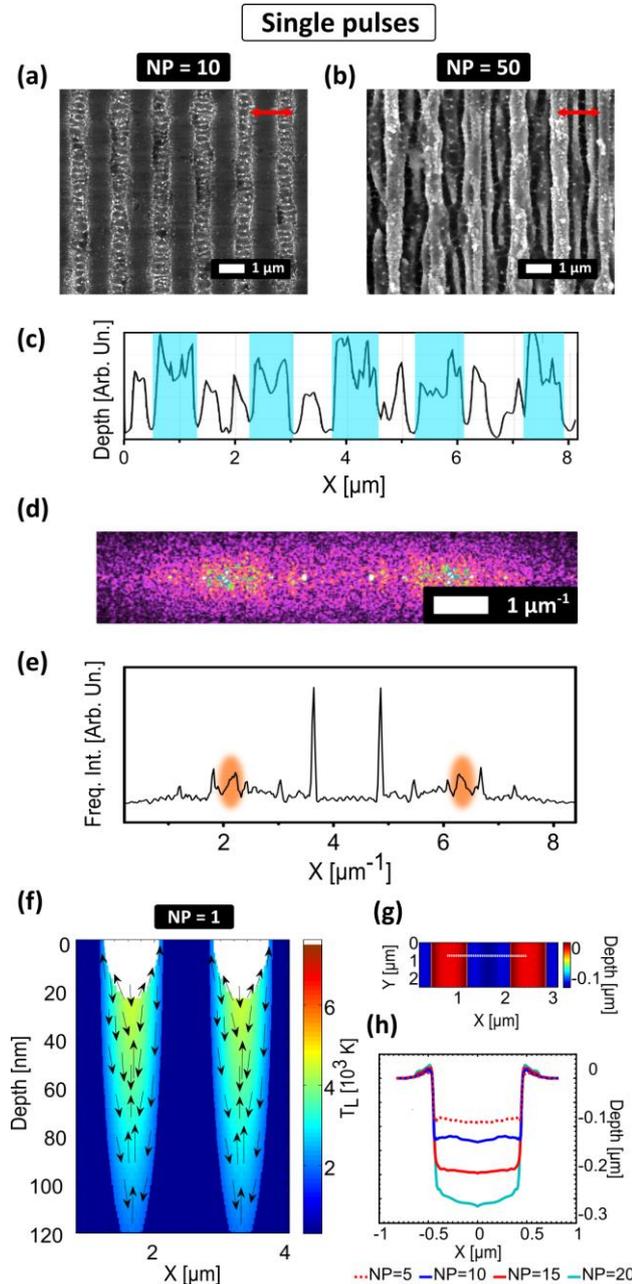

FIG. 3 Surface pattern upon *single* pulse irradiation with $\Lambda_{DLIP}$ ~1650 nm. SEM images of stainless steel surface for $NP$=10 (a) and $NP$=50 (b). (c) Cross section from (b). (d) FFT of (b) in a 30 x 30 µm² region (e) Cross section of (d.) (f) Modelling of the temperature profile and the flow vectors 450 ps after irradiation with first pulse. (g) Calculated surface profile after $NP$ = 10. (h). Depth profile along *white* dashed line in (g) for $NP$=5, 10, 15, 20. *Red* double-ended arrow in (a) and (b) indicates polarization direction.

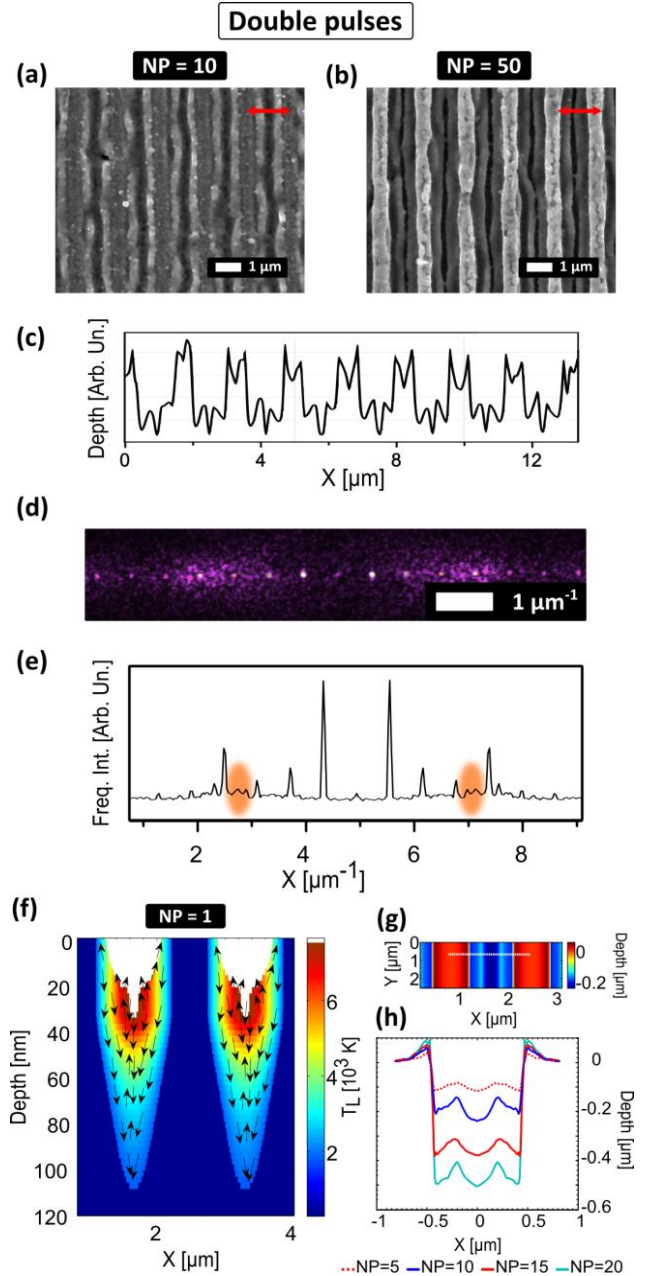

FIG. 4 Surface pattern upon *double* pulse irradiation with $\Lambda_{DLIP}$ ~1650 nm. SEM images of stainless steel surface for $NP$=10 (a) and $NP$=50 (b). (c) Cross section from b. (d) FFT of (b) in a 30 x 30 µm² region (e) Cross section of (d.) (f) Modelling of the temperature profile and the flow vectors 505 ps after irradiation with first pulse. (g) Calculated surface profile after $NP$ = 10. (h). Depth profile along *white* dashed line in (g) for $NP$=5, 10, 15, 20. *Red* double-ended arrow in (a) and (b) indicates polarization direction.

A similar methodology was followed to describe surface patterning for DP of a delay equal to $\tau_d$=500 ps. SEM images illustrate induced profiles for $NP$=10 ($\Lambda_{DLIP}$=1601 nm) (Fig.4a) and $NP$=50 ($\Lambda_{DLIP}$=1654 nm) (Fig.4b). The produced surface differs substantially from those of single pulses. At first, for $NP$ = 10, the surface morphology consists of a DLIP groove (Fig.4a). The HSFL structures observed for SP (Fig.3a) are not present in this case. This is possibly linked to the fact that the thermal effect as a result of the irradiation of a liquid with the second pulse of DP (i.e. stronger temperature gradients) might lead to weakening the electrodynamic phenomena that account for HSFL formation. Certainly, a consistent theory that predicts the first stages behind the formation of HSFL could also elucidate the development of elimination of those structures as a result of irradiation with DP.

Interestingly, for $NP$ = 50, the surface morphology consists of a very well ordered, periodic relief (Fig.4b) in contrast to the chaotic profile acquired for single pulses (Fig.3b). Again, the LIPSS observed for SP and $NP$ = 50 are not present here (Fig.3b). A cross section of Fig.4b is shown in Fig.4c. Furthermore, the FFT (Fig.4d) of a ~30 x 30 μm² area and its cross section across the polarization vector direction (Fig.4e) indicates a homogeneous structure formation. Comparing the FFT of SP and DP we note that the peak corresponding to LIPSS observed for SP (region in *orange* in Fig.3e) is not observed in the case of DP (region in *orange* in Fig.4e).

It is evident that the key role of the phase transition and the impact of irradiating a material in molten phase is revealed by simulations that accurately predict the obtained morphology. As explained in the previous section, a different thermal response of the material is expected for DP due to the fact that the second constituent pulse irradiates a part of the material in a liquid phase which is characterized by a distinct optical response; this leads further in different energy absorption and enhanced ablation which subsequently affects material reorganization [70]. Fig.4f illustrates the spatial temperature distribution at $t$=505 ps ($NP$=1) and the fluid movement. It is evident that due to the enhanced energy absorption, further fluid mass depression is produced at the centre of the crater where energy deposition is maximum which does not occur for single pulses. As a result, upon resolidification, a different corrugation profile is induced compared to the one due to single pulses (Fig.4g for $NP$=10, *blue* line in Fig.4h). The predicted value of the distance between the produced peaks is equal to ~0.41 μm while each peak is far from the DLIP highest position by ~0.30 μm that is comparable with the experimental value (~0.409 μm in Fig.4d). To illustrate the dynamic process that leads to the formation of the surface pattern with increasing $NP$, simulations results are presented in (Fig.4h) and in the Supplementary Material [70]). More specifically, the depth profile along the same place (*white* dashed line in Fig.4g) is illustrated in Fig.4h for $NP$=5, 10, 15, 20.

In conclusion, the difference in the structures obtained with SP and DP is emphasised and it can be attributed to the synergistic contribution of the electromagnetic coupling and absorption due to the distinct optical response between material in solid and liquid phase. Furthermore HSFL structures which are generally accepted to originate from near field effects [101] are observed only upon SP irradiation and are completely absent in the case of DP.

### b. Two laser beam DLIP with $\Lambda_{DLIP} \gg \Lambda_{LIPSS}$

A different structure pattern is developed for DLIP periods larger than $\lambda_L$ as SPP excitation can be achieved at those periods and yield structures big enough to support LSFL-SPP structures. Simulations for $\Lambda_{DLIP}$=5600 nm (experimental results were taken for $\Lambda_{DLIP}$ ~5276 μm (Fig.5a) and $\Lambda_{DLIP}$~5320 nm (Fig.5b)) indicate excitation of SPP modes, firstly, lead to the generation of periodic energy distribution as a result of the interference of the incident beam with the SPP waves and, secondly, yield LSFL structures *perpendicular* to LP with a calculated period equal to 680 nm for $NP$=10 assuming a computed SPP wavelength for the produced ripple height at $NP$=10 [21]. Fluid transport calculations assuming the attained temperature profiles (Fig.5c shows lattice temperature at $t$=450 ps) determine the final simulated profile (Fig.5c). The computed periodicity of the LSFL structures for $NP$=10 (Fig.5d) is relatively close to the experimental value (~608 nm). A profile of the surface corrugation with a DLIP period is illustrated in *blue* line in Fig.5e for $NP$=10. Similar results are attained for $NP$=50. Simulations appear to be in good agreement with experimental observations both qualitatively and quantitatively. To illustrate the dynamic process that leads to the formation of the surface pattern with increasing $NP$, simulations results are presented in (Fig.5e) and in the Supplementary Material [70]. More specifically, the depth profile along the same place (*white* dashed line in Fig.5d) is illustrated in Fig.5e for $NP$=5, 10, 15, 20.

As in the case of irradiation with single pulses, the employment of DP for ($\Lambda_{DLIP}$=5490 nm and $NP$=10) (Fig.6a) and larger ($\Lambda_{DLIP}$=5485 nm and $NP$=50 nm) (Fig.6b) yields periodic structure formation which is determined by the consideration of SPP excitation and fluid dynamical effects assuming the temperature variation induced by the application of the DLIP (Fig.6c). Simulations for ($\Lambda_{DLIP}$=5600 nm) (Fig.6c) yield periodicities equal to 687 nm for $NP$=10 which is a value close to the experimentally observed value ~714 nm (for $\Lambda_{DLIP}$=5490 nm and $NP$=10) and ~712 nm ($\Lambda_{DLIP}$=5485 nm and $NP$=50 nm). Furthermore, the surface pattern obtained with the use of SP (Fig.5c) is shallower compared to that with the employment of DP (Fig.6c) due to the additional depression of the crater surface following the irradiation of the molten material with the second pulse of DP. To illustrate the dynamic process that leads to the formation of the surface pattern with increasing *NP*, simulations results are presented in (Fig.6e) and in the Supplementary Material [70]. More specifically, the depth profile along the same place (*white* dashed line in Fig.6d) is illustrated in Fig.6e for *NP*=5, 10, 15, 20.

orientation is probably due to electrodynamic effects (i.e. impact of near field effects for low *NP*), however, more investigation is required to confirm the influence of electrodynamics.

By contrast, for *NP* = 50, the craters exhibit a rhombic shape and LIPSS are produced inside the craters with periods close to half of the laser wavelength. The measured period inside the craters varies from ~500 nm to ~750 nm. Modelling of the physical processes predict a surface

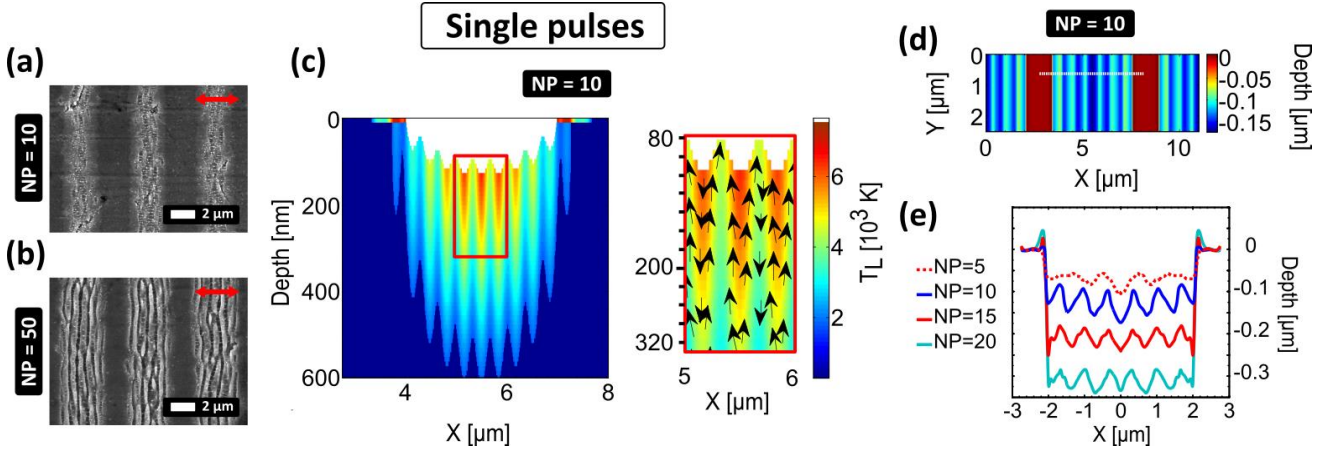

FIG. 5 Surface pattern upon *single* pulse irradiation with $\Lambda_{DLIP}$ ~5500 nm. SEM images of stainless steel surface for *NP*=10 (a) and *NP*=50 (b). (c) Modelling of the temperature profile and the flow vectors 450 ps after *NP* = 9. (d) Calculated surface profile after *NP* = 10. (e) Depth profile along *white* dashed line in (d) for *NP*=5, 10, 15, 20. *Red* double-ended arrow in (a) and (b) indicates polarization direction.

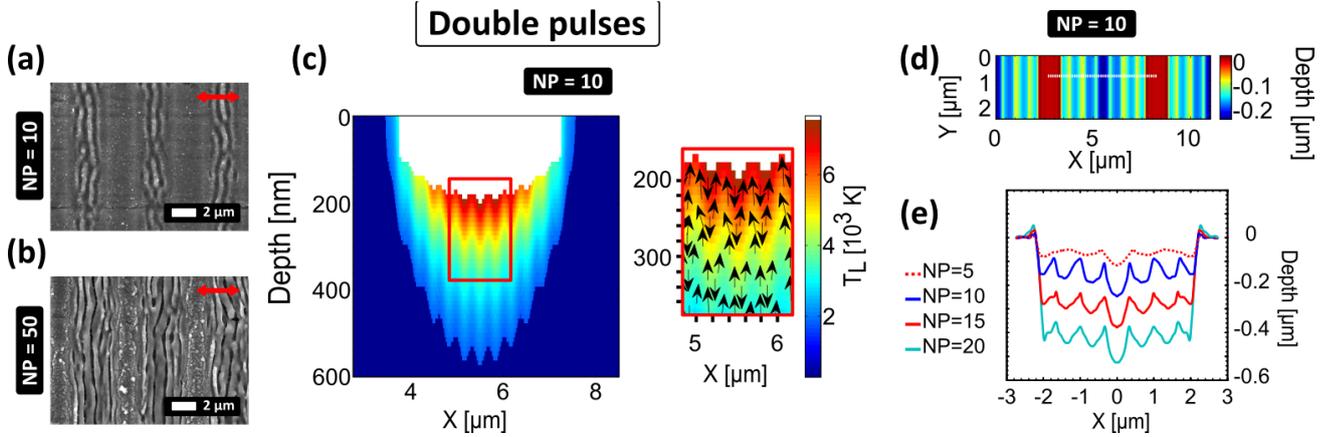

FIG. 6 Surface pattern upon *double* pulse irradiation with $\Lambda_{DLIP}$ ~5500 nm. SEM images of stainless steel surface for *NP*=10 (a) and *NP*=50 (b). (c) Modelling of the temperature profile and the flow vectors 505 ps after *NP* = 9. (d) Calculated surface profile after *NP* = 10. (e) Depth profile along *white* dashed line in (d) for *NP*=5, 10, 15, 20. *Red* double-ended arrow in (a) and (b) indicates polarization direction.

**c. Four laser beam DLIP (single and DP)**

While two beam- DLIP irradiation leads to formation of LSFL or HSFL structures, a four laser beam DLIP set up is expected to yield more complex structures due to the energy profile which is imposed on the material (Fig.2b). SP and DP are used with different $\Lambda_{DLIP}$ (~2262 nm or ~7600 nm) to determine the types of the induced patterns. For $\Lambda_{DLIP}$ ~2262 nm (Fig.7a-d), when *NP* = 10 a periodic array of craters decorated with HSFL is obtained. The long axis of the produced ellipsoidal shape is perpendicular to the laser polarization (Fig.7a,b). This preferential pattern in excellent agreement with experiment. Three lobes where developed inside the crater which is elongated parallel to the laser polarization direction (Fig.7e).

The elliptical shape of the craters with the long axis parallel to the laser polarisation in agreement to the simulation results and observations in previous reports for ULP [102]. FDTD calculations have revealed that the electric field distribution and, more specifically, local field enhancement effects in the direction of polarisation yields an elongation of either the crater or rippled areas along the polarisation vector for large *NP* or high fluences. In the present work, for which incorporation of near field effects are not considered, a different interpretation is presented. More specifically, in a four-beam DLIP technique, the direction of LP is perpendicular to the plane of incidence for two of the beams (*s*-polarised) while it is parallel to the plane of incidence for the other two (*p*-polarised). Due to the fact that reflectivity is higher (i.e. lower energy absorption) for *s*-polarised beams than for *p*-polarisation, a promotion of an elongation along LP is expected.

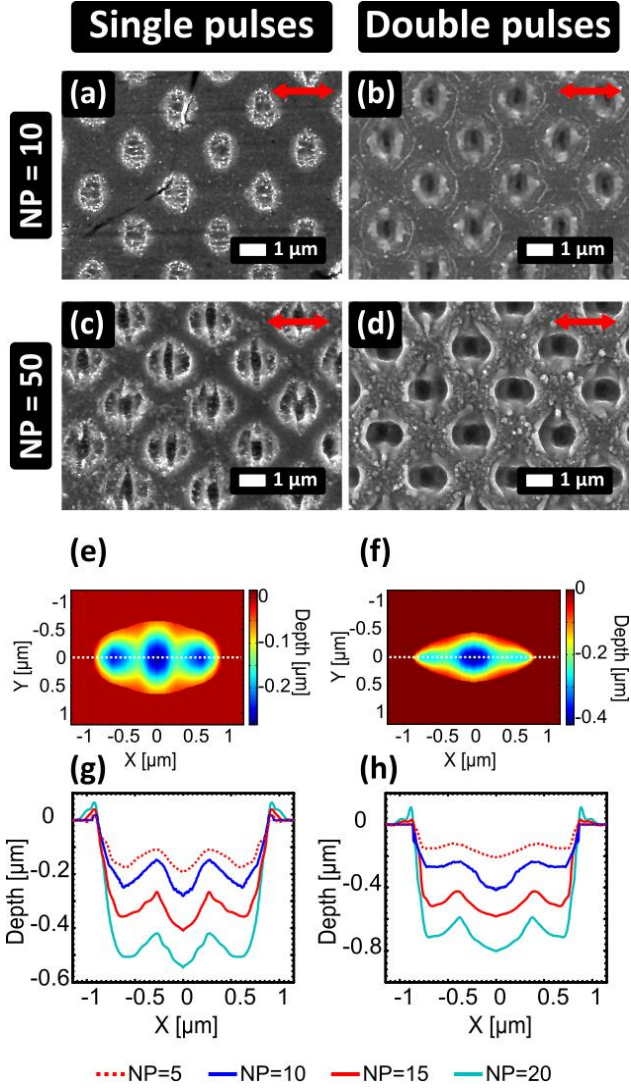
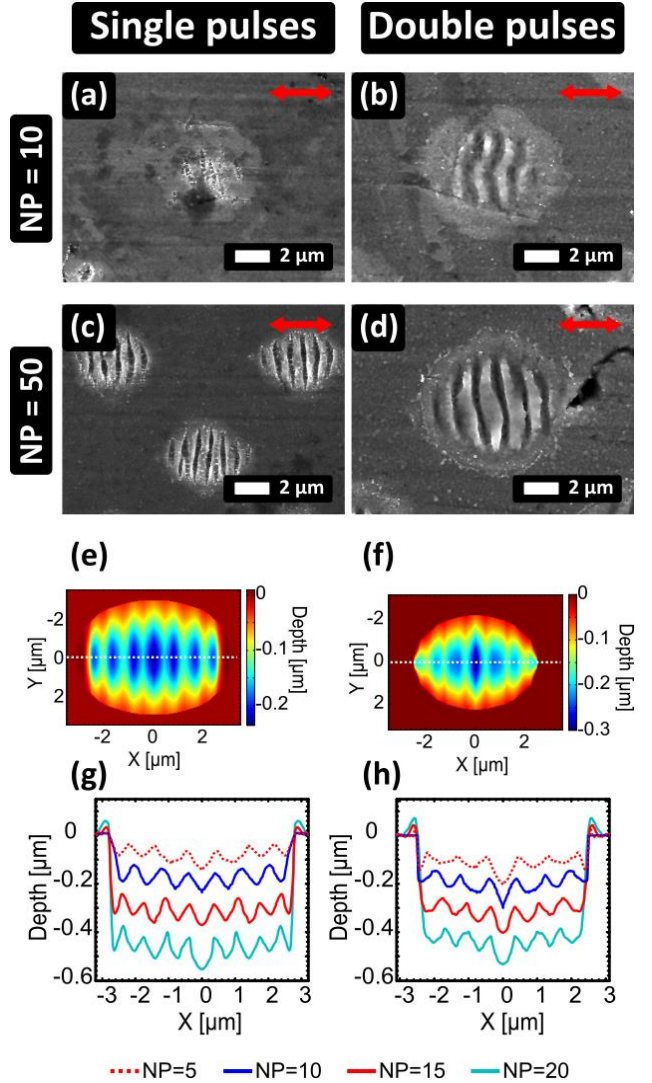

FIG. 7 SEM images of stainless steel surface irradiated with $Λ_{DLIP}$ ~2262 nm with single (a,c) and double pulses (b,d) with $NP$ = 10 and 50, respectively. Simulation results of the surface profile obtained with $NP$ = 10 are shown for SP (e) and DP. Depth profile along *white* dashed lines in (e) and (f) for $NP$=5, 10, 15, 20 are shown in (g) and (h).

FIG. 8 SEM images of stainless steel surface *irradiated* with $Λ_{DLIP}$ ~7600 nm with single (a,c) and double pulses (b,d) with $NP$ = 10 and 50, respectively. Simulation results of the surface profile obtained with $NP$ = 10 are shown for SP (e) and DP (f). Depth profile along *white* dashed lines in (e) and (f) for $NP$=5, 10, 15, 20 are shown in (g) and (h).

Irradiation with DP changes notably the obtained morphology (Fig.7b,d). At first, the when $NP$ = 10, we observe a crater which is elongated perpendicular to the laser polarization. Nonetheless, HSFL are not observed here as in all cases of DP irradiation. Interestingly, when $NP$ = 50, even if the conditions for SPP excitation are matched within the crater, LIPSS are not observed and the surface consists of an ellipsoidal crater elongated along the polarization vector with a round hole in the middle (Fig.7d). According to theoretical investigation, which reproduces accurately the experimentally obtained surface, the enhanced depression of the *crater* for DP (Fig.7f and [70]) leads to the disappearance of the lobes which are situated further from the crater centre. To illustrate the dynamic process that leads to the formation of the surface pattern with increasing $NP$ simulations results are presented Fig.7g-h and in the Supplementary Material [70]). More specifically, the depth profile along the same place (*white* dashed line in Fig.7e-f) is illustrated in Fig.7g-h, respectively, for $NP$=5, 10, 15, 20.

By contrast, for $Λ_{DLIP}$ ~7600 nm (Fig.8a-d) craters are formed on the surface. More specifically, for SP and $NP$=10 the craters are decorated with some HSFL structures (Fig.8a). For $NP$=50, a periodic pattern of craters with LSFL structures perpendicular to LP are formed (Fig.8c). Calculations have been performed to simulate LSFL production due to SPP-excitation and the predicted value is equal to 710 nm (Fig.8e) while the experimental values are in the range between 713 nm and 840 nm. On the other hand, upon DP irradiation, and $NP = 10$ LSFL are produced that agree with the experimental values (between ~840 nm and ~760 nm) for $NP = 50$ (Fig.8f). In general for $\Lambda_{DLIP} \gg \Lambda_{LIPSS}$ the surface morphology does not change significantly for either SP or DP apart from the universal observation that HSFL are not formed upon DP irradiation. To illustrate the dynamic process that leads to the formation of the surface pattern with increasing $NP$, simulations results are presented in Fig.8g-h and in the Supplementary Material [70]). More specifically, the depth profile along the same place (*white dashed line in Fig.8e-f*) is illustrated in Fig.8g-h, repectively, for $NP$=5, 10, 15, 20.

### d. Discussion of impact of DLIP and DP technique

The above investigation indicates that combining DLIP with DP enables the generation of novel morphologies in the near micron scale which demonstrates the capacity of the technique towards controlling laser induced morphology. On the other hand, the combined theoretical and experimental approach presented in this work aimed to set the basis for a description of the previously unexplored multiscale physical processes that lead to surface modification following the employment of DLIP with ULP for two- and four-beam. In particular, observations related to the interplay between the LIPSS and DLIP elucidates the structure formation mechanism; the emphatic impact of DP irradiation on the structure's morphology when $\Lambda_{DLIP} \sim \Lambda_{LIPSS}$ was demonstrated experimentally and interpreted theoretically. The fact that under confinement, the periods of LIPSS strongly depend on the DLIP indicates the common origin in the structure formation mechanism. This effect becomes more significant upon DP irradiation. On the other hand, simulations revealed the significant influence of ablation and hydrothermal effects in the formation of the laser-induced structures. Furthermore, both experimental observations and simulations showed that for a large pulse separation (~500 ps) between the constituent pulses of DP, LSFL and HSFL formation is suppressed when $\Lambda_{DLIP} \sim \Lambda_{LIPSS}$ both in 1D and 2D DLIP due to the particular microfluidic conditions; in that case, the morphology is dominated by the DLIP groove. Interestingly, irradiation with DP led to a distinct suppression in the crater where energy deposition is maximum which does not occur for single pulses. The theoretical predictions for the crater suppression (Fig.4f) and confirmed from the experimental observations highlighted also a very important aspect of laser irradiation that was not explored in the past, namely, the enhancement of energy absorption due to the distinct optical response of a material in a liquid phase.

In regard to the state-of-the-art of the modelling approach, the theoretical model was enriched with modules to account for ablation and simulate optical properties and energy absorption following irradiation of a material in molten phase; as stated above the theoretical simulations for DP and impact of irradiation of fluid with femtosecond pulses successfully describes the surface modifications. To the best of our knowledge, a similar theoretical approach to incorporate a module into the model that accounts for the response of a fluid material to fs irradiation has not been explored.

One interesting question that arises is whether all components/modules of the multiscale model are required to evaluate precisely the periodic structure formation and correlate the laser parameters with the induced morphology. It is evident that the answer is not straightforward; more specifically, the material type, its properties (i.e. optical or thermophysical) and the laser parameters can determine whether some approximations are applicable. For example, the laser conditions for the material used in this study showed that an abrupt drop of the reflectivity occurs at high fluences and temperatures which influences greatly the energy absorption and the thermal response of the irradiated solid; by contrast, in other materials or conditions, for which the optical response does not significantly varies, the calculation of transient reflectivity is unnecessary. Nevertheless, the theoretical framework developed and presented in this work, apart from addressing a realistic case, aims to constitute a complete approach and correlate the laser parameters and induced morphology for a general and not a specialized scenario.

While the production of most of the aforementioned structures were adequately predicted from the multiscale theoretical framework through modelling of the underlying physical processes, further revision of the theoretical framework is required to interpret the formation of deep subwavelength structures (HSFL). In addition, as noted above, a more precise estimate of the morphological features can be deduced by a more accurate evaluation of physical parameters at high temperatures (i.e. surface tension, recoil pressure, viscosity, density). It is evident that improving the control of the produced deterministic periodic textures with feature size down to the sub-micrometer range is expected to be important depending on the application (such as biological applications in terms of improved tribological, antibacterial and wetting properties [40-43]). Hence, an improved theoretical model can lead to a finer control of feature modulation.

Despite these limitations that can be the objective of a future work, the present study demonstrates the capability to control laser matter interaction through tailoring the coupling of DP and DLIP characteristic parameters (i.e. the interference period, polarization orientation, interpulse delay and number of incident pulses) to enable a novel surface engineering tool for advanced laser processing applications).

### VI. CONCLUSIONS

In this work, the combined action of DLIP and DP on laser induced structure formation has been investigated experimentally and theoretically. Results demonstrate the formation of well-ordered, novel morphologies with 1D and 2D symmetries while the predominant role of DLIP periodicity in the structure formation upon DP is also revealed. Furthermore, LIPSS formation within grooves with comparable size is severely impacted upon DP which emphasises the significance of the microfluidic reorganization of the surface. The fundamental physical mechanisms for the formation of tailored sub-micrometer periodic surface structures via tuning of the interplay between ultrashort-pulsed laser induced electrodynamics and melt hydrodynamics have been presented. One very important aspect that is revealed from the investigation of the physical mechanisms is associated to the irradiation of material in molten phase and the optical response of the fluid that influences both the fluid dynamics and induced surface pattern. The understanding of the underlying the mechanisms for DLIP coupled with ULP patterning is anticipated to shed light on novel laser-based processing techniques and identify routes for tailoring the morphology of a surface according to the demand of exciting applications, ranging from biomedical engineering to photovoltaics and nanoelectronics.

## ACKNOWLEDGEMENTS

The authors would like to acknowledge assistance of M. Vlachou and A. Lemonis in the experiments and A. Manousaki for the SEM characterization. Furthermore the authors acknowledge support by the European Union's Horizon 2020 research and innovation program through the project *BioCombs4Nanofibres* (grant agreement No. 862016). G.D.T and E.S. acknowledge funding from *HELLAS-CH* project (MIS 5002735), implemented under the "Action for Strengthening Research and Innovation Infrastructures" funded by the Operational Programme "Competitiveness, Entrepreneurship and Innovation" and co-financed by Greece and the EU (European Regional Development Fund) while G.D.T acknowledges financial support from COST Action *TUMIEE* (supported by COST-European Cooperation in Science and Technology).

Corresponding authors:
*Fotis Fraggelakis: fraggelakis@iesl.forth.gr
♣ George D.Tsibidis: tsibidis@iesl.forth.gr (Theory and Simulations)
♦Emmanuel Stratakis: stratak@iesl.forth.gr